# The scaling laws of human travel

D. Brockmann[*], L. Hufnagel[‡] & T. Geisel[*]

[*]*Max-Planck Institute for Dynamics and Self-Organisation, Bunsenstr. 10, 37073 Göttingen and Department of Physics, University of Göttingen, Germany*

[‡]*Kavli Institute for Theoretical Physics, University of California, Santa Barbara, CA 93106, USA*

**The dynamic spatial redistribution of individuals is a key driving force of various spatiotemporal phenomena on geographical scales. It can synchronise populations of interacting species, stabilise them, and diversify gene pools[1-3]. Human travelling, e.g. is responsible for the geographical spread of human infectious disease[4-9]. In the light of increasing international trade, intensified human mobility and an imminent influenza A epidemic[10] the knowledge of dynamical and statistical properties of human travel is thus of fundamental importance. Despite its crucial role, a quantitative assessment of these properties on geographical scales remains elusive and the assumption that humans disperse diffusively still prevails in models. Here we report on a solid and quantitative assessment of human travelling statistics by analysing the circulation of bank notes in the United States. Based on a comprehensive dataset of over a million individual displacements we find that dispersal is anomalous in two ways. First, the distribution of travelling distances decays as a power law, indicating that trajectories of bank notes are reminiscent of scale free random walks known as Lévy flights. Secondly, the probability of remaining in a small, spatially confined region for a time T is dominated by algebraically long tails which attenuate the superdiffusive spread. We show that human travelling behaviour can be described mathematically on many spatiotemporal scales by a two parameter continuous time random walk model to a surprising accuracy and conclude that human travel on geographical scales is an ambivalent effectively superdiffusive process.**

The quantitative aspects of dispersal in ecology are based on the dispersal curve which quantifies the relative frequency of travel distances of individuals as a function of geographical distance[11,12]. A large class of dispersal curves (e.g. exponential, Gaussian, stretched exponential) permits the identification of a typical length scale by the variance of the displacement length or equivalent quantities. When interpreted as the probability $p(r)$ of finding a displacement of length r in a short time $\delta t$ the existence of a typical length scale often justifies the description of dispersal in terms of diffusion equations on large spatiotemporal scales. If, however, $p(r)$ lacks a typical length scale, i.e. $p(r) \sim r^{-(1+\beta)}$ with $\beta < 2$, the diffusion approximation fails. In physics, random processes with such a single step distribution are known as Lévy flights[13-17] (see supplementary information). Recently, the related notion of long-distance-dispersal (LDD) has been established in dispersal ecology[18], taking into account the observation that dispersal curves of a number of species exhibit tails due to long range movements[19-22].

Nowadays, humans travel on many spatial scales, ranging from a few to thousands of kilometres in short periods of time. The direct quantitative assessment of human movements, however, is difficult and a statistically reliable estimate of human dispersal comprising all spatial scales does not exist. The central idea of this work is to utilise data collected at online bill tracking websites which monitor the worldwide dispersal of large numbers of individual bank notes and to infer the statistical properties of human dispersal with very high spatiotemporal precision. Our analysis of human movement is based on trajectories of 464,670 dollar bills obtained from the bill tracking system www.wheresgeorge.com. We analysed the dispersal of bank notes in the United States, excluding Alaska and Hawaii. The core data consists of 1,033,095 reports to the bill tracking website. From these reports we calculated the geographical displacements $r = |\mathbf{x}_2 - \mathbf{x}_1|$ between a first ($\mathbf{x}_1$) and secondary ($\mathbf{x}_2$) report location of a bank note and the elapsed time $T$ between successive reports.

In order to illustrate qualitative features of bank note trajectories, Fig. 1b depicts short time trajectories ($T < 14$ days) originating from three major cities (Seattle, WA, New York, NY, Jacksonville, FL). Succeeding their initial entry, the majority of bank notes are reported next in the vicinity of the initial entry location, i.e. $|\mathbf{x}_2 - \mathbf{x}_1| \lesssim 10\,\mathrm{km}$ (Seattle: 52.7%, New York: 57.7% Jacksonville: 71.4%). However, a small yet considerable fraction is reported beyond a distance of 800 km (Seattle: 7.8%, New York: 7.4%, Jacksonville: 2.9%).

From a total of 20,540 short time trajectories originating all across the United States we measured the probability $p(r)$ of traversing a distance $r$ in a time interval $\delta T$ between one and four days (Fig. 1c). A total of 14,730 (i.e. a fraction $Q = 0.71$) secondary reports occurred outside a short range radius $L_{min} = 10$ km. Between $L_{min}$ and the approximate average east-west extension of the United States $L_{max} \approx 3,200$ km, the kernel exhibits power law behaviour $p(r) \sim r^{-(1+\beta)}$ with an exponent $\beta = 0.59 \pm 0.02$. For $r < L_{min}$, $p(r)$ increases linearly with r which implies that displacements are distributed uniformly inside the disk $|\mathbf{x}_2 - \mathbf{x}_1| \lesssim L_{min}$. We measured $p(r)$ for three classes of initial entry locations: highly populated metropolitan areas (191 sites, local population: $N_{loc} > 120,000$), cities of intermediate size (1,544 sites, local population: $120,000 > N_{loc} > 22,000$), and small towns (23,640 sites, local population: $N_{loc} < 22,000$) comprising 35.7 %, 29.1 % and 25.2 % of the entire population of the United States, respectively. The inset of Fig. 1c depicts $p(r)$ for these classes. Despite systematic deviations for short distances, all distributions exhibit an algebraic tail with the same exponent $\beta \approx 0.6$, which confirms that the observed power-law is an intrinsic and universal property of dispersal.

However, the situation is more complex. If we assume that the dispersal of bank notes can be described by a Lévy flight with a short time probability distribution $p(r)$, we can estimate the time $T_{eq}$ for an initially localised ensemble of bank notes to reach the stationary dis-tribution[23] (maps in Fig. 1a) and obtain $T_{eq} \approx 68$ days (see supplementary information). Thus, after 2-3 months bank notes should have reached the equilibrium distribution. Surprisingly, the long time dispersal data does not reflect a relaxation within this time. Fig. 1b shows secondary reports of bank notes with initial entry at Omaha, NE which have dispersed for times $T > 100$ days (with an average time $\langle T \rangle = 289$ days). Only 23.6% of the bank notes travelled farther than 800 km, the majority of 57.3% travelled an intermediate distance $50 < r < 800$ km and a relatively large fraction of 19.1% remained within a radius of 50 km even after an average time of nearly one year. From the computed value $T_{eq} \approx 68$ days a much higher fraction of bills is expected to reach the metropolitan areas of the West Coast and the New England states after this time. This is sufficient evidence that the simple Lévy flight picture for dispersal is incomplete. What causes this attenuation of the dispersal?

Two alternative explanations might account for this effect. The slowing down might be caused by the strong spatial inhomogeneity of the system. People might be less likely to leave large cities than e.g. suburban areas. Alternatively, long periods of rest might be an intrinsic temporal property of dispersal. In as much as an algebraic tail in the spatial displacements yields superdiffusive behaviour, a tail in the probability density $\phi(t)$ for times t between successive spatial displacements of an ordinary random walk can lead to subdiffusion[16] (see supplementary information). Here, the ambivalence between scale free spatial displacements and scale free periods of rest can be responsible for the observed attenuation of superdiffusion.

In order to address this issue we investigated the relative proportion $P_0^i(t)$ of bank notes which are reported in a small (20 km) radius of the initial entry location i as a function of time (Fig. 1d). The quantity $P_0^i(t)$ estimates the probability for a bank note of being reported at the initial location at time t. We computed $P_0^i(t)$ for metropolitan areas, cities of intermediate size and small towns: For all classes we found the asymptotic behaviour $P_0(t) \sim At^{-\eta}$ with the same exponent $\eta = 0.6 \pm 0.03$ which indicates that waiting time and dispersal characteristics are universal. Notice that for a pure Lévy flight with index $\beta$ in two dimensions $P_0(t)$ scales with time as $t^{-2/\beta}$ (dashed red line)[16]. For $\beta \approx 0.6$ (as suggested by Fig. 1c) this implies $\eta \approx 3.33$, i.e. a five fold steeper decrease than observed which clearly shows that dispersal cannot be described by a pure Lévy flight. The measured decay is even slower than the decay expected from ordinary two-dimensional diffusion ($\eta = 1$, dashed black line). Therefore, we conclude that the slow decay in $P_0(t)$ reflects the impact of an algebraic tail in the distribution of rests $\phi(t)$ between displacements. Indeed, if $\phi(t) \sim t^{-(1+\alpha)}$ with $\alpha < 1$, then $\eta = \alpha$ and consequently $\alpha = 0.60 \pm 0.03$. This suggests that an algebraic tail in the distribution of rests $\phi(t)$ is responsible for slowing down the superdiffusive dispersal advanced by the short time dispersal kernel of Fig. 1c.

In order to model the antagonistic interplay between scale free displacements and waiting times we use the framework of continuous time random walks (CTRW) introduced by Montroll and Weiss[24]. A CTRW consists of a succession of random displacements $\delta\mathbf{x}_n$ and random waiting times $\delta t_n$ each of which is drawn from a corresponding probability density function $p(\delta\mathbf{x}_n)$ and $\phi(\delta t)$. After N iterations the position of the walker and the elapsed time are given by $\mathbf{x}_N = \sum_n \delta\mathbf{x}_n$ and $t_N = \sum_n \delta t_n$. The quantity of interest is the position $\mathbf{x}(t)$ after time t and the associated probability density $W(\mathbf{x},t)$ which can be computed within CTRW-theory. For displacements with *finite* variance $\sigma^2$ and waiting times with *finite* mean $\tau$ such a CTRW yields *ordinary* diffusion asymptotically, i.e. $\partial_t W(\mathbf{x},t) = D\partial_x^2 W(\mathbf{x},t)$ with a diffusion coefficient $D = \sigma^2 / \tau$.

In contrast, we assume here that both, $p(\delta\mathbf{x}_n)$ and $\phi(\delta t)$ exhibit algebraic tails, i.e. $p(\delta\mathbf{x}_n) \sim |\delta\mathbf{x}_n|^{-(1+\beta)}$ and $\phi(\delta t) \sim |\delta t|^{-(1+\alpha)}$, for which $\sigma^2$ and $\tau$ are *infinite*. In this case we can derive a bifractional diffusion equation for the dynamics of $W(\mathbf{x},t)$:

$$\partial_t^\alpha W(\mathbf{x},t) = D_{\alpha,\beta}\partial_{|x|}^\beta W(\mathbf{x},t). \tag{1}$$

In Eq. 1 the symbols $\partial_t^\alpha$ and $\partial_{|x|}^\beta$ denote fractional derivatives which are non-local and depend on the tail exponents $\alpha$ and $\beta$. The constant $D_{\alpha,\beta}$ is a generalised diffusion coefficient (see supplementary information). Eq. 1 represents the core dynamical equation of our model. Using methods of fractional calculus we can solve this equation and obtain the probability $W_r(r,t)$ of having traversed a distance r at time t,

$$W_r(r,t) = t^{-\alpha/\beta}L_{\alpha,\beta}\left(r/t^{\alpha/\beta}\right), \tag{2}$$

where $L_{\alpha,\beta}$ is a universal scaling function which represents the characteristics of the process. Eq. 2 implies that the typical distance travelled scales according to $r(t) \sim t^{1/\mu}$ where $\mu = \beta/\alpha$. Thus, depending on the ratio of spatial and temporal exponents the random walk can be effectively either superdiffusive ($\beta < 2\alpha$), subdiffusive ($\beta > 2\alpha$), or quasidiffusive ($\beta = 2\alpha$) (see supplementary information). For the exponents observed in the dispersal data ($\beta = 0.59 \pm 0.02$ and $\alpha = 0.60 \pm 0.03$) the theory predicts a temporal scaling exponent in the vicinity of unity, $\mu = 0.98 \pm 0.08$. Therefore, dispersal remains superdiffusive despite long periods of rest.

The validity of our model can be tested by estimating $W_r(r,t)$ from the entire dataset of a little over half a million displacements and elapse times. The scaling property is best extracted from the data by a transformation to logarithmic coordinates $z = \log_{10} r$, $\tau = \log_{10} t$ and the associated probability density $W_z(z, \tau)$. If the original process scales according to $r(t) \sim t^{1/\mu}$, the density $W_z(z, \tau)$ is a function of $z - \tau / \mu$ only. Fig. 2a shows that scaling occurs in a time window between approximately 7 days and one year. From the slope (blue line) we obtain a scaling exponent $\mu = 1.05 \pm 0.02$ which agrees well with our model.

Finally, we investigated to which degree bank note dispersal exhibits a scaling density as predicted by our model, i.e. relation 2. Fig. 2b shows $t^{1/\mu} W_r(r, t)$ extracted from data versus the ratio $y = r / t^{1/\mu}$. The exponent $\mu = 1.05$ was set to the value obtained in Fig. 2a. The data collapse on a single curve indicates that in the chosen time interval of 10 to 365 days bank note dispersal exhibits a universal scaling function. The asymptotics of the empirical curve is given by $y^{-(1-\xi_1)}$ and $y^{-(1+\xi_2)}$ for small and large arguments, respectively. Both exponents fulfil $\xi_1 \approx \xi_2 \approx 0.6$. We compared the empirical curve with the theoretical prediction of our model. By series expansions one can compute the asymptotics of the limiting function $L_{\alpha,\beta}(y)$ in Eq. 2 which gives $y^{-(1-\beta)}$ and $y^{-(1+\beta)}$ for small and large y, respectively. Consequently, as $\beta \approx 0.6$ (Fig. 1c) the theory agrees well will the observed exponents. For the entire range of y we computed $L_{\alpha,\beta}(y)$ by numeric integration for $\alpha = \beta = 0.6$ and superimposed the theoretical curve on the empirical one. The agreement is very satisfactory. In summary, our analysis gives solid evidence that the dispersal of bank notes can be accounted for by our model.

The question remains how the dispersal characteristics of bank notes carry over to the travelling behaviour of humans. In this context one can conclude that the power law with exponent $\beta = 0.6$ of the short time dispersal kernel for bank notes reflects the human dispersal kernel because the exponent remains unchanged for short time intervals of T=2,4,7 and 14 days. The issue of long waiting times is more subtle. One might speculate that the observed algebraic tail in waiting times of bank notes is a property of bank note dispersal alone. Long waiting times may be caused by bank notes which exit the money tracking system for a long time, for instance in banks. However, if this were the case the inter-report time statistics would exhibit an algebraic tail as well. Analysing the inter-report time distribution we found an exponential decay which suggests that bank notes are passed from person to person at a constant rate. If we assume that humans exit small areas at a constant rate which is equivalent to exponentially distributed waiting times and that bank notes pass from person to person at a constant rate, the distribution of bank note waiting times would also be exponential in contrast to the observed power law. This reasoning permits in our eyes no other conclusion than a lack of scale in human waiting time statistics. We obtained further support for our results from a comparison with two independent human travelling datasets: long distance travel on the United States aviation network[8, 25, 26] and the latest survey on long distance travel conducted by the United States Bureau of Transportation Statis-tics[27] (see supplementary information). Both agree well with our findings and support our conclusions.

Based on our analysis we conclude that the dispersal of bank notes and human travelling behaviour can be described by a continuous time random walk process incorporating scale free jumps as well as long waiting times between displacements. To our knowledge this is the first empirical evidence for such an ambivalent process in nature. We believe that our results will serve as a starting point for the development of a novel class of models for the spread of human infectious diseases, because universal features of human travel can now be accounted for in a quantitative way.

**Acknowledgements:** We would like to thank the initiators of the bill tracking system www.wheresgeorge.com and hope this work will contribute to its popularity. We thank cabinet maker Dennis Derryberry for lively discussions and for drawing our attention to the wheresgeorge website. We thank Boris Shraiman, Doron Cohen, and W. Noyes for reading the manuscript and their valuable comments.

**Author contributions**: Idea: D.B. & L.H., data pre-processing: L.H., data analysis: D.B. & L.H., theory and model: D.B., manuscript: D.B., L.H. & T.G.

**Author Information:** The authors declare no competing financial interests. Correspondence and requests for materials should be addressed to D.B. (e-mail: zwerg@chaos.gwdg.de).

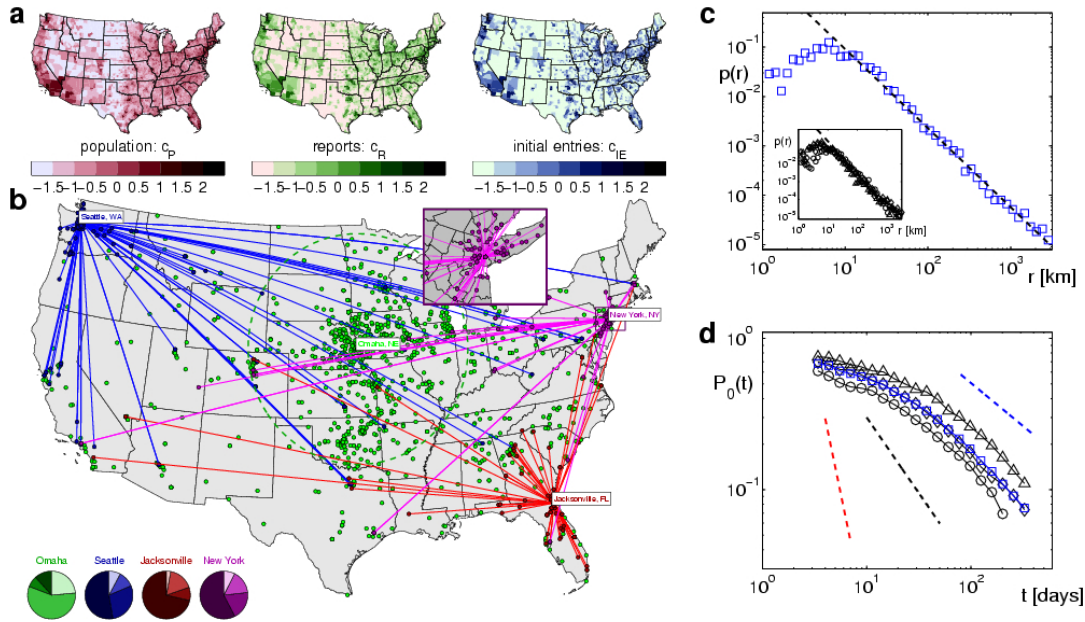

**Figure 1| Dispersal of bank notes and humans on geographical scales. a,** Relative logarithmic population, report and initial entry densities $c_P = \log \rho_P / \langle \rho_P \rangle$, $c_R = \log \rho_R / \langle \rho_R \rangle$ and $c_{IE} = \log \rho_{IE} / \langle \rho_{IE} \rangle$ as functions of geographical coordinates. Colour encodes the densities relative to the nation-wide averages (3,109 counties) $\langle \rho_P \rangle = 95.15$, $\langle \rho_R \rangle = 0.34$ and $\langle \rho_{IE} \rangle = 0.15$ of individuals, reports and initial entries per $km^2$, respectively. **b,** Trajectories of bank notes originating from four different places. Tags indicate initial, symbols secondary report locations. Lines represent short time trajectories with travelling time $T < 14$ days. Lines are omitted for the long time trajectories (initial entry: Omaha) with $T > 100$ days. The inset depicts a close-up of the New York area. Pie charts indicate the relative number of secondary reports coarsely sorted by distance. The fractions of secondary reports that occurred at the initial entry location (dark), at short ($0 < r < 50\,km$), intermediate ($50 < r < 800\,km$) and long ($r > 800\,km$) distances are ordered by increasing brightness of hue. The total number of initial entries are $N = 2,055$ (Omaha), $N = 524$ (Seattle), $N = 231$ (New York), $N = 381$ (Jacksonville). **c,** The short time dispersal kernel. The measured probability density function $p(r)$ of traversing a distance $r$ in less than $T = 4$ days is depicted by blue symbols. It is computed from an ensemble of 20,540 short time displacements. The dashed black line indicates a power law $p(r) \sim r^{-(1+\beta)}$ with an exponent of $\beta = 0.59$. The inset depicts $p(r)$ for three classes of initial entry locations (black triangles for metropolitan areas, diamonds for cities of intermediate size, and circles for small towns). Their decay is consistent with the measured exponent $\beta = 0.59$ (dashed line). **d,** The relative proportion $P_0(t)$ of secondary reports within a short radius ($r_0 = 20\,km$) of the initial entry location as a function of time. Blue squares depict $P_0(t)$ averaged over 25,375 initial entry locations. Black triangles, diamonds, and circles show $P_0(t)$ for the same classes as in **c**. All curves decrease asymptotically as $t^{-\eta}$ with an exponent $\eta = 0.60 \pm 0.03$ indicated by the blue dashed line. Ordinary diffusion in two dimensions predicts an exponent $\eta = 1.0$ (black dashed line). Lévy flight dispersal with an exponent $\beta = 0.6$ as suggested by **b** predicts an even steeper decrease, $\eta = 3.33$ (red dashed line).

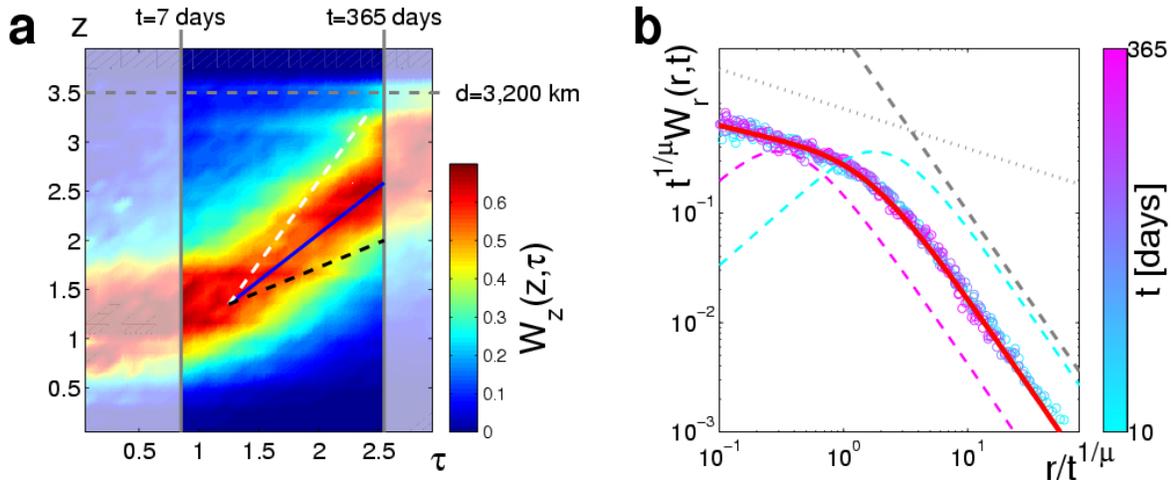

**Figure 2 | Spatiotemporal scaling of bank note dispersal. a**, the probability density $W_z(z, \tau)$ of having travelled a logarithmic distance $z = \log_{10} r$ at logarithmic time $\tau = \log_{10} t$. The middle segment indicates the scaling regime between one week and one year. The superimposed red line represents the scaling behaviour $r(t) \sim t^{1/\mu}$ with exponent $\mu = 1.05 \pm 0.05$. It is compared to the diffusive scaling (black dashed line) and the scaling of a pure Lévy process with exponent $\beta = 0.6$ (white dashed line). The upper dashed grey line shows the approximate linear extent $L_{max} = 3,200\,km$ of the United States. **b**, The measured radial probability density $W_r(r, t)$ and theoretical scaling function $L_{\alpha, \beta}(r/t^{1/\mu})$ (Eq. 2). In order to extract the quality of scaling the function $t^{1/\mu} W_r(r, t)$ is shown for various but fixed values of $t$ between 10 and 365 days as a function of the rescaled distance $r/t^{1/\mu}$, where the exponent $\mu$ was set to the value determined in **a**. As the measured (circles) curves collapse on a single curve the process exhibits universal scaling. The scaling curve represents the limiting density of the process. The asymptotic behaviour for small (grey dotted line) and large (grey dashed line) arguments $y = r/t^{1/\mu}$ is given by $y^{-(1-\xi_1)}$ and $y^{-(1+\xi_2)}$, respectively, with estimated exponents $\xi_1 = 0.63 \pm 0.04$ and $\xi_1 = 0.62 \pm 0.02$. According to our model these exponents must fulfil $\xi_1 = \xi_2 = \beta$ where $\beta$ is the exponent of the asymptotic short time dispersal kernel (Fig. 1c), i.e. $\beta \approx 0.6$. The superimposed red line represents $t^{1/\mu} W_r(r, t)$ predicted by our theory with spatial and temporal exponents $\alpha = 0.6$ and $\beta = 0.6$, respectively. The coloured dashed lines represent $t^{1/\mu} W_r(r, t)$ for a pure Lévy flight with $\beta = 0.6$ at the times $t = 10$ and $t = 365$ days. The curves do not collapse because the pure Lévy flight exhibits the wrong spatio-temporal scaling. Furthermore, the limiting curves strongly deviate from the data for small arguments.